\documentclass[amsmath,11pt,amssymb,aps]{revtex4}

\setcitestyle{super}

\usepackage{color}
\usepackage[]{graphicx}
\usepackage{latexsym}
\usepackage{natbib}
\usepackage{amsmath}
\usepackage[caption=false]{subfig}
\usepackage{multirow}

\begin{document}

\title{High-resolution of particle contacts via fluorophore exclusion
  in deep-imaging of jammed colloidal packings}

\author{Eru Kyeyune-Nyombi$^{1,2}$, Flaviano Morone$^{1,3}$, Wenwei
  Liu$^{1,4}$, Shuiqing Li$^4$, M. Lane
  Gilchrist$^{2}$ \footnote{Corresponding author: gilchrist@ccny.cuny.edu}, and Hern\'an
  A. Makse$^{1,3,}$ \footnote{Corresponding author: hmakse@lev.ccny.cuny.edu}}

\affiliation{$^1$ Levich Institute, City College of New York, New
  York, NY 10031, USA \\ $^2$ Department of Chemical Engineering and
  Department of Biomedical Engineering, City College of New York, New
  York, NY 10031, USA \\ $^3$ Department of Physics, City College of
  New York, New York, NY 10031, USA \\ $^4$ Department of Thermal Engineering, Tsinghua University, Beijing 100084, China
}

\begin{abstract}
{\bf Understanding the structural properties of random packings of
  jammed colloids requires an unprecedented
  high-resolution determination of the contact network providing
  mechanical stability to the packing. Here, we address the
  determination of the contact network by a novel strategy based on
  fluorophore signal exclusion of quantum dot nanoparticles from the
  contact points. We use fluorescence labeling schemes on particles
  inspired by biology and biointerface science in conjunction with
  fluorophore exclusion at the contact region. The method provides
  high-resolution contact network data that allows us to measure
  structural properties of the colloidal packing near marginal
  stability. We determine scaling laws of force distributions, soft
  modes, correlation functions, coordination number and free volume
  that define the universality class of jammed colloidal packings and
  can be compared with theoretical predictions. The
  contact detection method opens up further experimental testing at
  the interface of jamming and glass physics. }

\end{abstract}

\maketitle
\clearpage

\section{I\MakeLowercase{ntroduction}}

The problem with the experimental investigation of 
jammed colloidal systems \cite{jaeger,solomon,baule} 
is that it is difficult to look inside
of a particulate packing. This is especially problematic
from a theoretical standpoint. While recent
theoretical advances have provided a fresh perspective on the long-standing
packing problem --- including replica theory from spin glasses,
constraint satisfaction problems, geometrical and force ensembles ---
most of these theories are built from the bottom up
\cite{edwards2,vanhecke,forcemap,kurchan,swm,torquato,zamponi1,wyart1,wyart2,charbonneau}.
Therefore, experimentally testing these theories requires full
information of the contact network at sufficiently high resolution for
resolving fragile contacts at the state of marginal stability observed
during the jamming transition. 
Theoretical predictions of observables like coordination number
(number of contacting particles), the scaling of the small-force
distribution and geometrical order parameters require exact
determination of contact between any two particles.  While methods,
like X-ray tomography, help analyze contacts between large grains
\cite{aste1,aste2} (having diameter sizes on the order of mm) these
methods are not as effective for studying jammed matter on
colloidal length scales. Resolution is limited even in deep imaging of dense jammed colloids
\cite{weeks2} and jammed emulsions
\cite{brujic,jorjadze,dinsmore,brujic2,clusel} using confocal microscopy.
Better resolution is needed if one seeks to determine whether two
colloidal particles are in contact at a fragile state of marginal stability
near the jamming transition. Here we use the
fluorophore signal exclusion of quantum dot nanoparticles at the
contact points to determine the contact network with higher
resolution \cite{NyombiThesis}.

\begin{figure}[h]
\centering{
(a)\includegraphics[width=5cm]{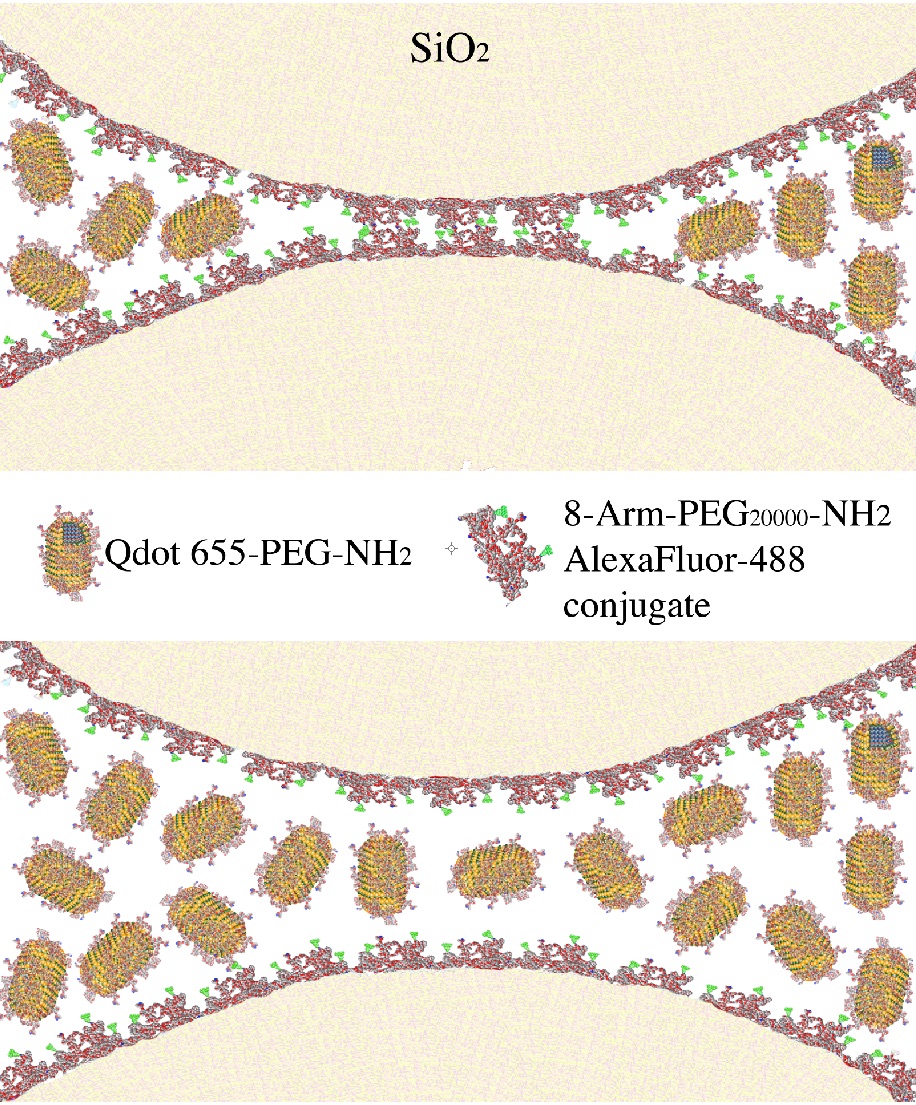} 
(b)\includegraphics[width=10cm]{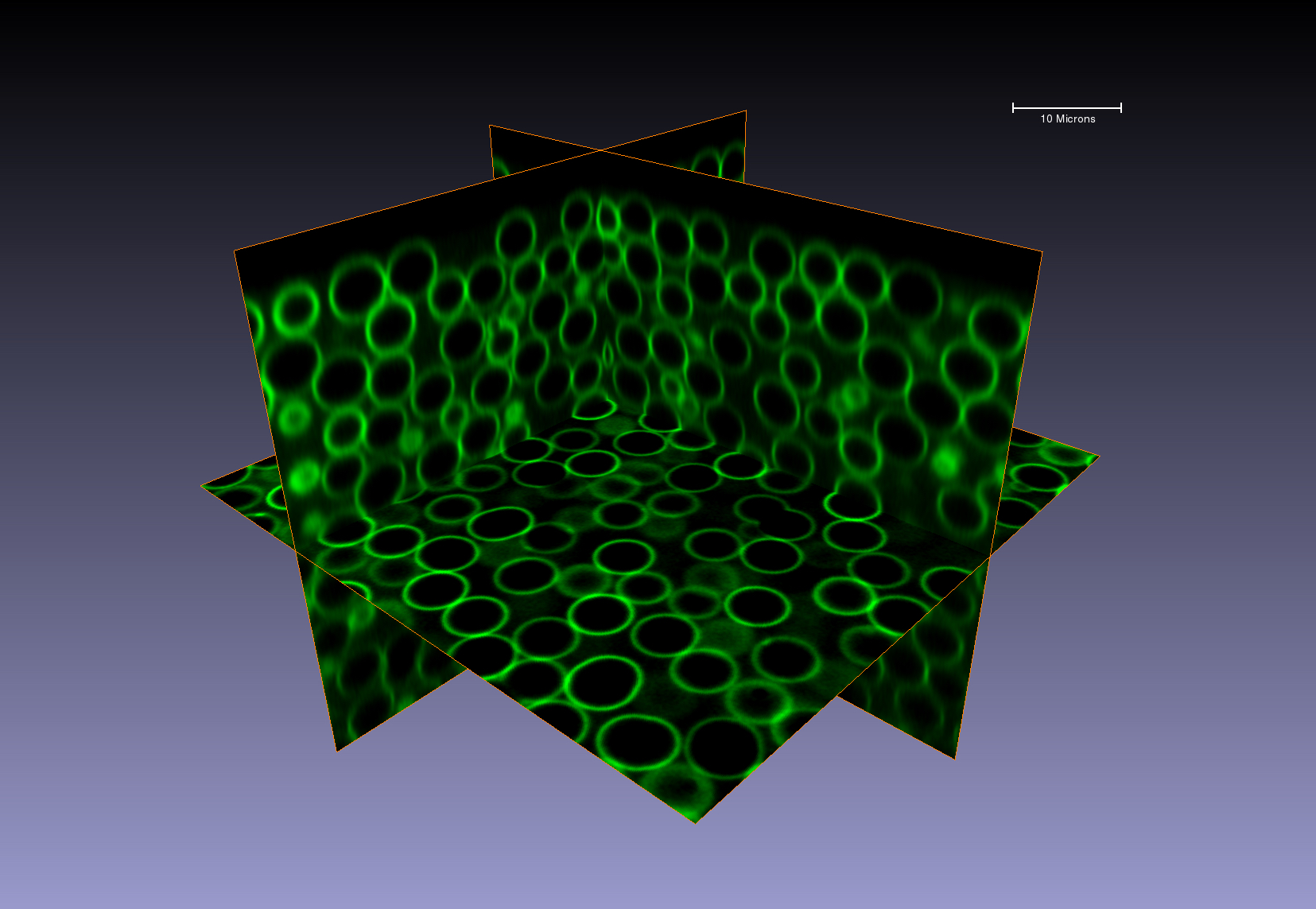}\\
(c)\includegraphics[width=10cm]{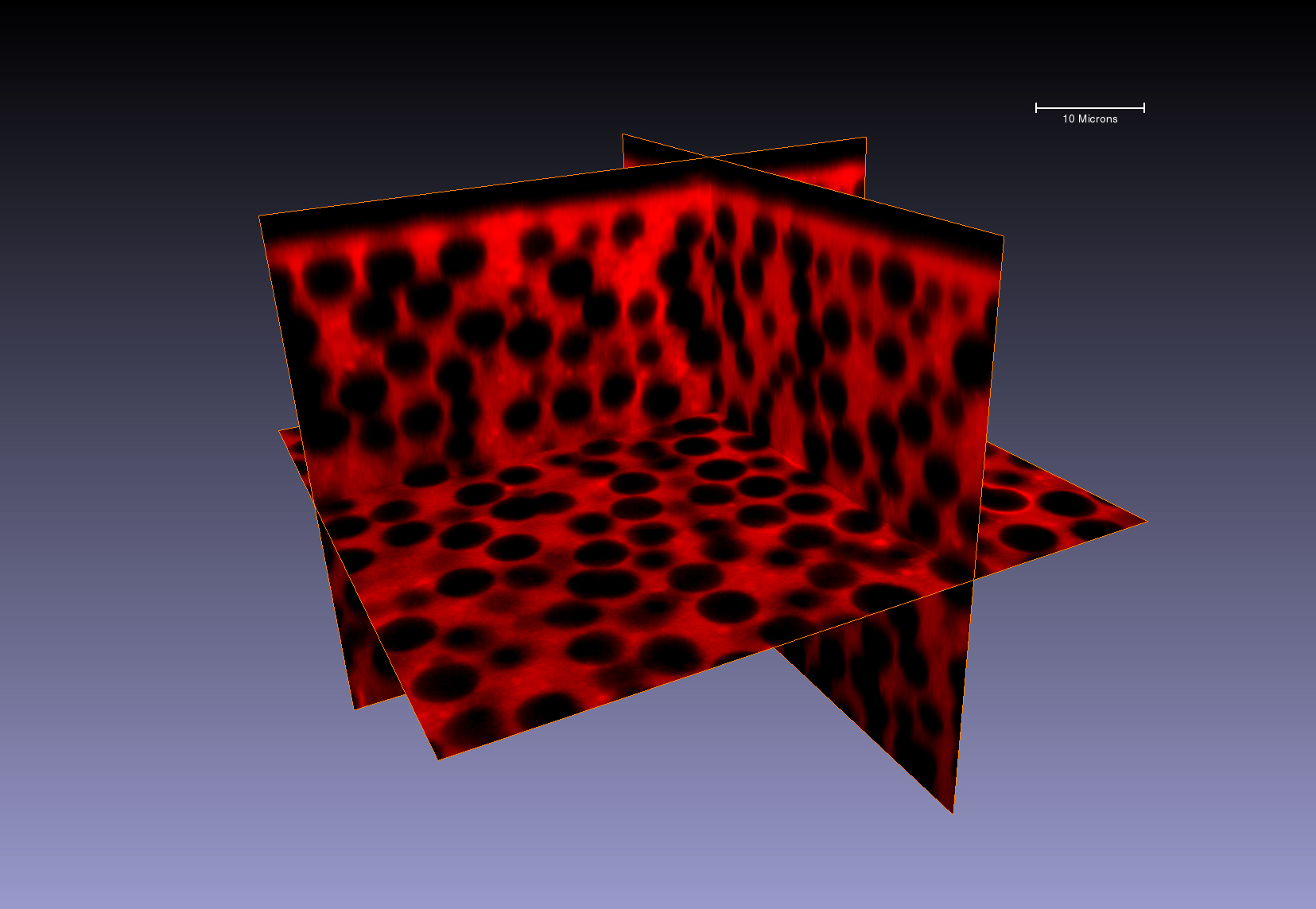}
}
\caption{(a) Schematic of contact gap between particles.
AlexaFluor\textsuperscript{\textregistered} 488 (AF)
is attached to the surface of the particles
(8-arm-PEG$_{20,000}$-AlexaFluor488).  
QD nanoparticles (Qdot655-PEG-NH$_2$) are in solution.  
QD nanoparticles are excluded from the contact gap 
in the top panel but they occupy the contact gap in 
the bottom panel.
(b) 3D confocal image of the packing.  Only showing 
green fluorescence from AF on particles' surface 
(i.e., the fluorescent `rings'), as previously 
demonstrated \cite{brujic,dinsmore,brujic2,clusel}.
(c) Same image of the packing but only showing 
red fluorescence from QD nanoparticles in solution.}
\label{fig:schematic}
\end{figure}

\section{E\MakeLowercase{xperimental methods}}

\subsection{S\MakeLowercase{ample Preparation}}



We consider a colloidal system of green  
fluorescent (em: 515--555nm) silica microspheres
in an aqueous-glycerol solution containing red fluorescent (em: 600--700nm)
quantum dot (QD) nanoparticles.
The green fluorescence on the surface of the microspheres comes from 
AlexaFluor\textsuperscript{\textregistered} 488 (AF) manufactured by Life
Technologies, Inc.  This fluorescent dye is attached to surface of silica 
microspheres using amine and N-hydroxysuccinimide (NHS) ester reaction 
chemistry\cite{Hermanson}.  First Bis(succinimidyl) nona(ethlyene glycol), 
manufactured by ThermoFisher Scientific, Inc. is conjugated 
onto the surface of $5.06\pm 0.44{\mu}$m SiO$_{2}$-NH$_{2}$ microspheres, 
manufactured by Bang Laboratories. The same chemistry then attaches 8-Arm 
polyethylene glycol (PEG) star-polymer (MW = 20kDa), manufactured by
Nanocs, Inc. and finally attaches 
AlexaFluor\textsuperscript{\textregistered} 488, AF 
(emission: 515--555nm).  This surface chemistry is necessary for uniform 
fluorescence that is devoid of dark patches upon the surface of each 
particle.  The added PEG also 
produces short-range steric repulsive interactions \cite{Israelachvilli} 
that largely negates adhesive forces and minimizes
friction between particles \cite{Drobek}.  This allows 
particles to move freely and to pack randomly in aqueous-glycerol solution.

The size and density of the fluorescent silica particles 
causes gravitational settling to exceed Brownian motion 
(i.e., P{\'e}clet number $\sim 329$).  Hence, the particles 
pack naturally by sedimentation. Therefore, the ``pressure'' 
in each packing equals the weight of the particles themselves.
Centrifugation momentarily adds to the weight or pressure of
each packing to insure compaction.  No centrifugation was 
used on packing A listed in table \ref{tab:exponents}.  
Packings B and C were each centrifuged at 2210$\times g$
and 4416$\times g$, respectively, 
using an Eppendorf centrifuge (model 5804R).

Quantum dot (QD) nanoparticles manufactured by Thermo 
Fisher Scientific, Inc. (cylindrically shaped and approximately 
$8\times15$ nm in size\cite{Giepmans} with a narrow red emission peak 
at 600--700nm) are added to the aqueous-glycerol 
solution of the compacted particulate system.  
The refractive index of the solution is carefully designed to 
match the refractive index of the particles which allows for 
deep-imaging into the packing using confocal microscopy.  
Fig. \ref{fig:schematic}b-c  show two 3D confocal 
images of green and red fluorescence from the \textit{same} packing 
acquired \textit{at the same time} using different fluorescent channels.

\begin{figure}[h]
\centering{
(a)\includegraphics[width=12cm]{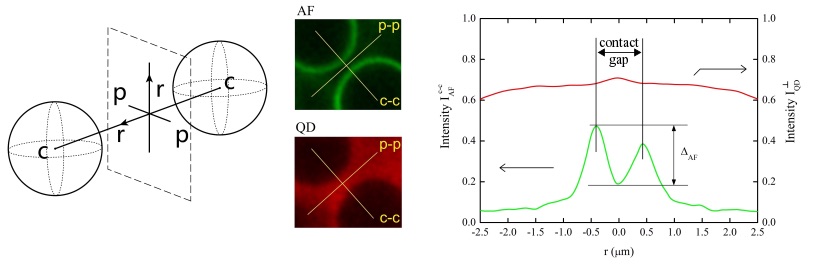}\\
(b)\includegraphics[width=10cm]{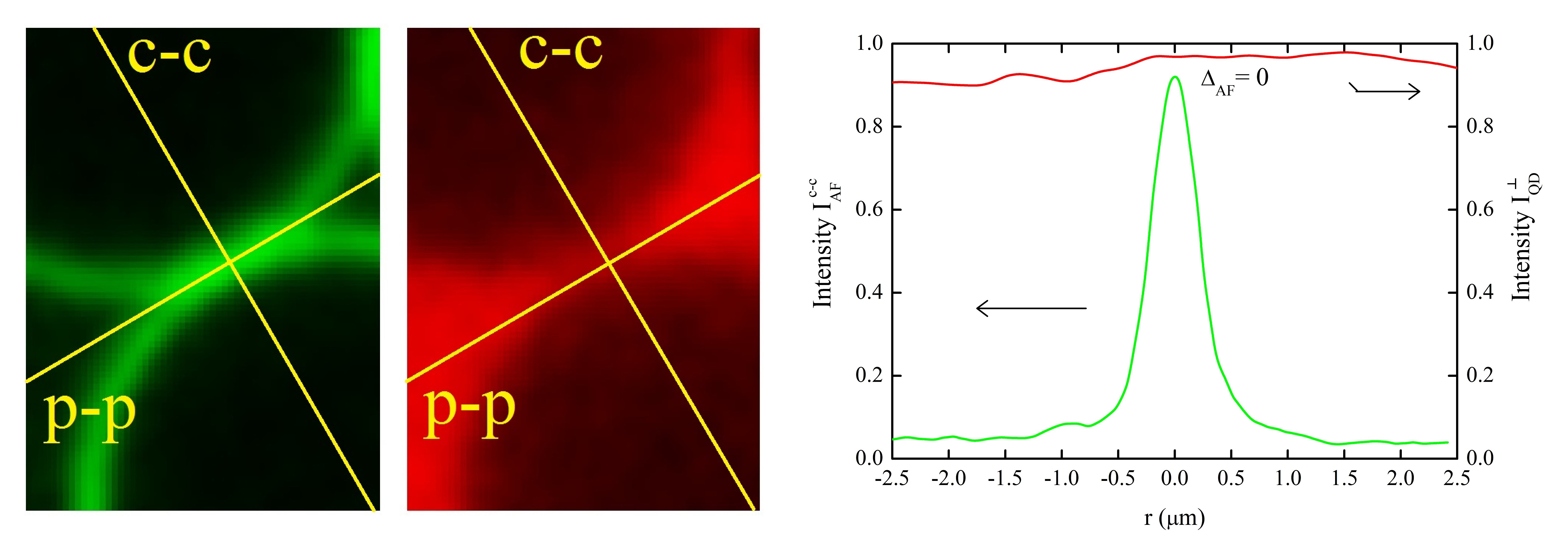}\\
(c)\includegraphics[width=10cm]{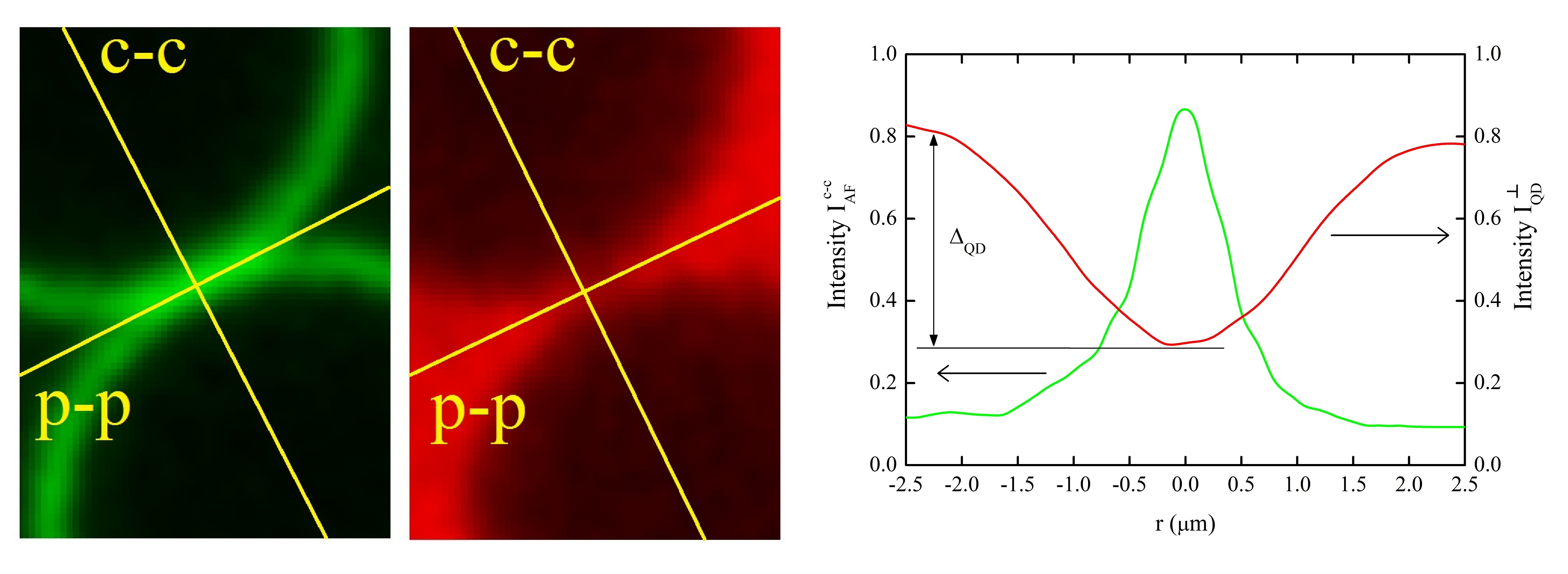}
}
\caption{Measurement of intensity profiles in AF,
$I_{\rm AF}^{\rm c-c}$, and QD, $I^{\perp}_{\rm QD}$,
along lines c-c and p-p, respectively.  
Edges of particles identified 
by peaks in $I_{\rm AF}^{\rm c-c}$.  
(a) Two separate distinct peaks in
$I_{\rm AF}^{\rm c-c}$. Therefore, contact gap $>$ 230nm.
(b) Convolution of two peaks into a single peak in 
$I_{\rm AF}^{\rm c-c}$ (i.e., $\Delta_{\rm AF} = 0$).
Therefore contact gap $<$ 230nm.  However, flat profile 
in $I^{\perp}_{\rm QD}$ suggests no physical exclusion 
of QDs. Hence, non-contact or false positive contact. 
(c)  Again, two peaks convolve into a single peak
in $I_{\rm AF}^{\rm c-c}$ (i.e., $\Delta_{\rm AF} = 0$). 
Therefore contact gap $<$ 230nm.  But dip in 
$I^{\perp}_{\rm QD}$, which is measured by 
$\Delta_{\rm QD}$, suggests physical exclusion of QDs.
Hence, probable contact.}
\label{fig:contact}
\end{figure}

\subsection{A\MakeLowercase{nalysis}}

The contact detection method probes the exclusion of quantum dot
nanoparticles (QDs) from the contact gap between any
two neighboring colloidal particles and also monitors the
emission of AF at the surface of the particles. 
Fig. \ref{fig:schematic}a provides a schematic representation 
of QDs being excluded from the contact gap. 
The exclusion of QDs is accompanied by a noticeable decrease 
in fluorescent signal from QDs.  This decreasing signal intensity 
correlates with the size of the inter-particle
space from which the QDs are excluded and thus provides a measure
of the contact gap \cite{Asakura} in a process that
we call {\it fluorophore signal exclusion}.
We note that, technically, we are not measuring the size of QDs as in
super-resolution Stimulated-Emission-Depletion (STED) techniques
\cite{Hell,Betzig2}. Thus, our detection method does not break the
diffraction limit.  Instead, we infer
information on the contact gap by measuring the fluorescent signal
from QDs and correlating it with the gap between particles.

Fig. ~\ref{fig:contact}a-c exemplify,
for three characteristic cases of contacts, the intensity profile of
AF ($I^{\rm c-c}_{\rm AF}$) along a line c-c between the
centers of nearest neighboring particles and the
intensity profile of QDs ($I^{\perp}_{\rm QD}$) along a line p-p that is
perpendicular to the line c-c as sketched in
Fig. \ref{fig:contact}a, left panel. 

Fig. \ref{fig:contact}a shows the case of well-separated
particles. In this case, two adjacent peaks in $I_{\rm AF}^{\rm c-c}$
locate the edges of neighboring particles (the green rings of
AF). These peaks represent the bounding edges of the
inter-particle space or the contact gap.  
Incidentally, the distance between the peaks 
is also the size of the contact gap.  The flat profile of
$I^{\perp}_{\rm QD}$ observed in
Fig. \ref{fig:contact}a, right panel, represents QDs inside the
contact gap that have not experienced exclusion due to the 
large size of the gap.  
In this case, the absence of a contact is
unambiguous.  We characterize similar cases of well-separated 
particles by measuring image-contrast, $\Delta_{\rm AF}$, 
which is defined as the difference between the
highest and lowest values in $I^{\rm c-c}_{\rm AF}$ 
between the two peaks (see Fig. \ref{fig:contact}a, right panel).  
When $\Delta_{\rm AF} > 0$, the two
peaks in $I_{\rm AF}^{\rm c-c}$ are clearly distinguishable 
and the size of the contact gap is well above the diffraction 
limit.  However, as $\Delta_{\rm AF} \rightarrow 0^+$ the two peaks
convolve into a single peak and the size of 
the contact gap becomes unresolvable.

\begin{figure}
\includegraphics[width=.75\textwidth]{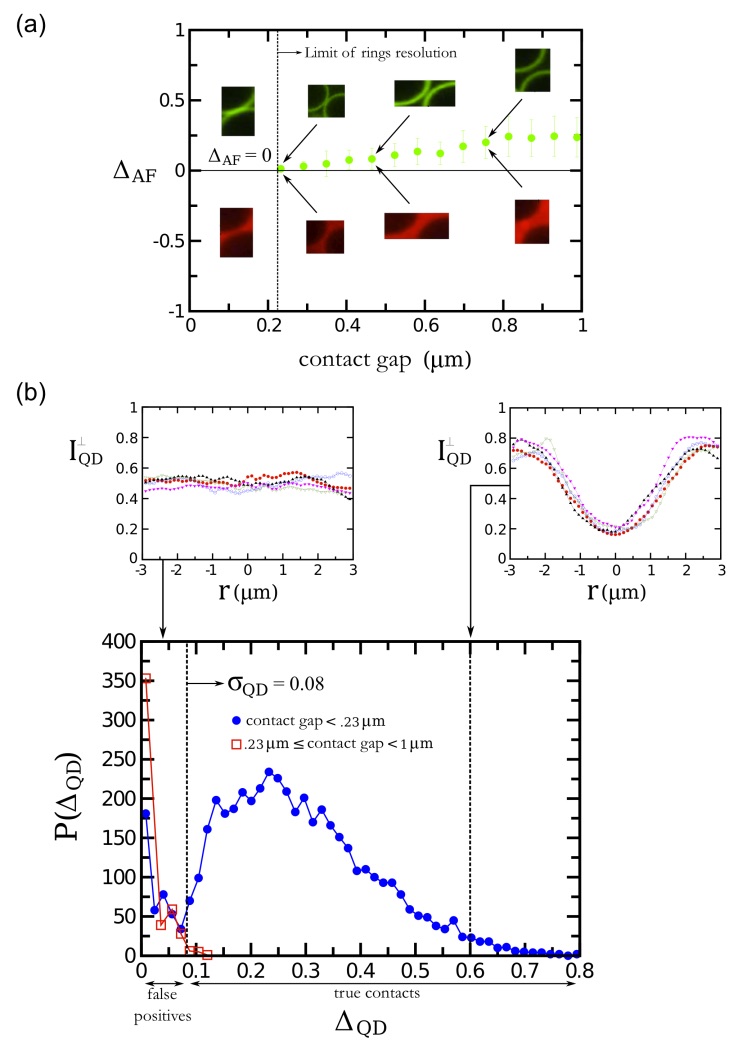} 
\caption{(a) Average values from numerous measurements
in image contrast, $\Delta_{\rm AF}$, for each 
contact gap size.  Error bars denote stand. dev. 
in measurements of $\Delta_{\rm AF}$ for any given 
contact gap size.  Fluorescent rings are indistinguishable 
(i.e., $\Delta_{\rm AF} = 0$) and the contact gap 
appears to vanish below the diffraction limit 
at 0.23$\mu$m.  (b) $\Delta_{\rm QD}$ measures 
the extent of the dip in $I^{\perp}_{\rm QD}$ (top two panels)
and monitors the likelihood that a contact gap has vanished 
for particles to touch. Distribution 
of dips in QD fluorescence, $P(\Delta_{\rm QD})$, for 
contact gaps smaller than 0.23$\mu$m (bottom panel).
Dips in $I^{\perp}_{\rm QD}$ where 
$\Delta_{\rm QD} < \sigma_{\rm QD}$ exhibit random 
fluctuations (top left panel) whereas larger 
values, like $\Delta_{\rm QD}= 0.6$, show more
distinct fluorophore signal exclusion (top right panel) 
and stronger evidence of contact.}
\label{fig:fig3} 
\end{figure}

In the case shown in Fig. \ref{fig:contact}b, the AF-detected contact
gap disappears, $\Delta_{\rm AF} = 0$, and neighboring particles
seemingly touch.  However, the emergence of a single peak in $I^{\rm
  c-c}_{\rm AF}$ does not necessarily mean that a contact has been
established. It simply means that 
the AF rings cannot be resolved any further. 
This resolution-limit occurs at 0.23$\mu$m
(see Fig. \ref{fig:fig3}a). Although $\Delta_{\rm AF} = 0$ 
and particles appear to be touching in 
Fig. \ref{fig:contact}b, we still observe a flat profile
in $I^{\perp}_{\rm QD}$ similar to that seen
in Fig. \ref{fig:contact}a.  The flat profile in 
$I^{\perp}_{\rm QD}$ suggests the contact gap is still large 
enough to accommodate QDs.  Therefore, even though
the rings have merged, we identify contacts like Fig. \ref{fig:contact}b 
as non-contacts or false positive contacts.  
This improves upon earlier methods \cite{brujic,dinsmore,brujic2,clusel}, 
where monitoring the fluorescence from 
AF rings alone would have identified similar non-contacts as
positive contacts.  

Now instead of relying upon AF rings alone, a dip in 
$I^{\perp}_{\rm QD}$, resolves 
the contact gap at higher resolution
to identify the most probable positive contacts.
Fig. \ref{fig:contact}c shows a dip in $I^{\perp}_{\rm QD}$ 
that is measured as $\Delta_{\rm QD}$.
The signature of a resolved contact at higher resolution is the
emergence of a dip in QD fluorescence signal exclusion as seen in
Fig. \ref{fig:contact}c. This occurs when $\Delta_{\rm AF} = 0$
and the contact gap is smaller than 0.23$\mu$m, as illustrated by
Fig. \ref{fig:fig3}a which plots the average value of 
$\Delta_{\rm AF}$ for numerous pairs of particles at
different sizes of contact gaps. 
0.23$\mu$m marks the maximum resolution of the contact gap using AF rings.
A higher resolution of the contact gap is achieved by 
monitoring the emergence of
a dip in $I^{\perp}_{\rm QD}$ using $\Delta_{\rm QD}$.  

Fig. ~\ref{fig:fig3}b shows the distribution of dips in 
QD fluorescence signal exclusion, $P(\Delta_{\rm QD})$,
for different contact gaps.  
For particles with contact gaps smaller than 230nm 
(see blue circles), Fig. ~\ref{fig:fig3}b
shows two populations of contacts well separated by a minimum in
$P(\Delta_{\rm QD})$. 
We also notice that the minimum in $P(\Delta_{\rm QD})$ appears for
a value of $\Delta_{\rm QD}$ equal to the average standard
deviation, $\sigma_{\rm QD}$, of $I^{\perp}_{\rm QD}$ for flat profiles. 
This minimum in $P(\Delta_{\rm QD})$ is observed at 
$\Delta_{\rm QD} \approx \sigma_{\rm QD} = 0.08$. 
Therefore, dips in $I^{\perp}_{\rm QD}$ 
having $\Delta_{\rm QD} < 0.08$ 
may be considered as random fluctuations since they are
indistinguishable from the random fluctuations in the profile itself
(i.e., they fall below 1$\sigma$).  This observation introduces
a natural definition of a false positive contact for
contact gaps smaller than 230nm.  
After plotting $P(\Delta_{\rm QD})$ for contact gaps
ranging from 230nm to 1$\mu$m (see red squares),
we find that $P(\Delta_{\rm QD})$ peaks at 
$\Delta_{\rm QD}=0$ and falls off quickly
to zero when $\Delta_{\rm QD} \approx \sigma_{\rm QD}$. This result
suggests that our method of contact detection is able to
find false positives that previous methods could
not detect. 

In summary, our method measures the contact gap as the distance
between peaks in 
$I_{\rm AF}^{\rm c-c}$, as seen in Fig. \ref{fig:contact}a.
When these peaks merge below 230nm (as seen in Fig. 
\ref{fig:contact}b-c) it is not possible to know the 
exact size of the contact gap for these contacts.  
It is at this point that we analyze the QD dip profile 
to differentiate false positives from more probable contacts.

\section{R\MakeLowercase{esults}}

Next, we employ our contact detection method to determine the contact
network of jammed colloidal packings.  Theoretical findings
posit a common physical explanation for jamming and glass transitions 
~\cite{kurchan,zamponi1,charbonneau,mari,degiuli,linbo,wyart1,wyart2}. The
emerging unifying laws are condensed in a set of key observables
characterized by critical scaling exponents.  But experimentally testing theoretical models
requires high resolution of contacts to
differentiate fragile contacts from close (non-contacting) neighbors. These fragile contacts
contribute greatly to the marginal stability of jammed colloids
which materializes into scaling laws for small forces, soft modes and
near-contact pair correlation function
\cite{kurchan,zamponi1,charbonneau,mari,degiuli,linbo,lin,wyart1,wyart2}. The
contact detection method outlined here determines the contact network at high resolution
and presents a unique opportunity for assessing key physical observables.

We analyze colloidal packings $A, B$ and $C$ (Table
\ref{tab:exponents}) and for each of them we measure the
inter-particle force distribution, the pair correlation function, the
vibrational density of states and the equation of state relating
coordination number and volume fraction.
We have jammed the packings at different centrifugation to
study variations in properties with respect to the preparation protocol.
The experimentally obtained packing fractions and coordination
numbers are close enough, so that we are not able to reach definitive
conclusions regarding protocol dependence.

The forces at the contacts can be calculated solely from the contact
network as determined by the fluorophore exclusion technique without
resorting to a contact force law, by using the fact that the system is
close to isostatic.  We resolve the contact forces at the contact
points using the force network ensemble method proposed
earlier~\cite{vanhecke} and calculate the force distribution as
follows:
\begin{itemize}
\item (i) The force balance equations are imposed as constraints:
\begin{equation}
\sum_{a\in\partial i}\vec{f}^i_a = 0\ \ \ \ \ \ i=1,\dots, N\ ,
\end{equation}
where the notation $a\in\partial i$ indicates the set of contact points 
$a$ around particle $i$;
\item(ii) Forces are repulsive, i.e.:
\begin{equation}
\vec{f}^i_a\cdot\vec{d}^i_a < 0\ , 
\end{equation}
where we denoted by $\vec{d}^i_a$ the vector connecting 
the $i$th particle’s position (of its center of mass) $r_i$
and the $a$th contact on the $i$th particle;
\item(iii) A fixed external force $P$ sets an overall force scale.
\end{itemize}

Then, we compute the force distribution $P_{a}(f)$ at
the contact points $a$ by using a simulated annealing 
algorithm. In practice, we use a penalty function given by equation \ref{penaltyFunct} which disfavors
force configurations that do not satisfy conditions $(i)-(iii)$ above.

\begin{equation}
E = \sum_{a}\left(\sum_{j\in\partial a}\vec{f}^i_a\right)^2 + 
\left(\sum_{a}\sum_{j\in\partial a}|\vec{f}^i_a| - P\right)^2
\label{penaltyFunct}
\end{equation}

Finally, we obtain the mean force distribution 
$P(f)=\langle P_a(f)\rangle$ by averaging over all the contacts.

Theoretically, at the jamming transition, the force distribution
$P(f)$ is expected to decay algebraically for $f\to 0^+$ as seen in
equation \ref{forceEqn}:

\begin{equation}
P(f)\sim f^{\min(\theta,\theta')}
\label{forceEqn}
\end{equation} 

where the exponents $\theta$ and $\theta'$ describe, respectively,
localized and delocalized
excitations~\cite{wyart1,wyart2,charbonneau,zamponi1,linbo}.  Mean
field theory of hard sphere glasses in infinite
dimensions~\cite{charbonneau} predicts only the value of the exponent
$\theta'\sim0.42311$, since in infinite dimensions, where the mean
field theory has been developed, there are no localized
excitations. To adhere as much as possible to the theory, in this work
we measure only the exponent $\theta'$ by excluding localized
excitations as done elsewhere~\cite{buckling}. This is done by
removing the so-called bucklers.  

\begin{figure}
\centerline{(a)\includegraphics[width=.5\textwidth]{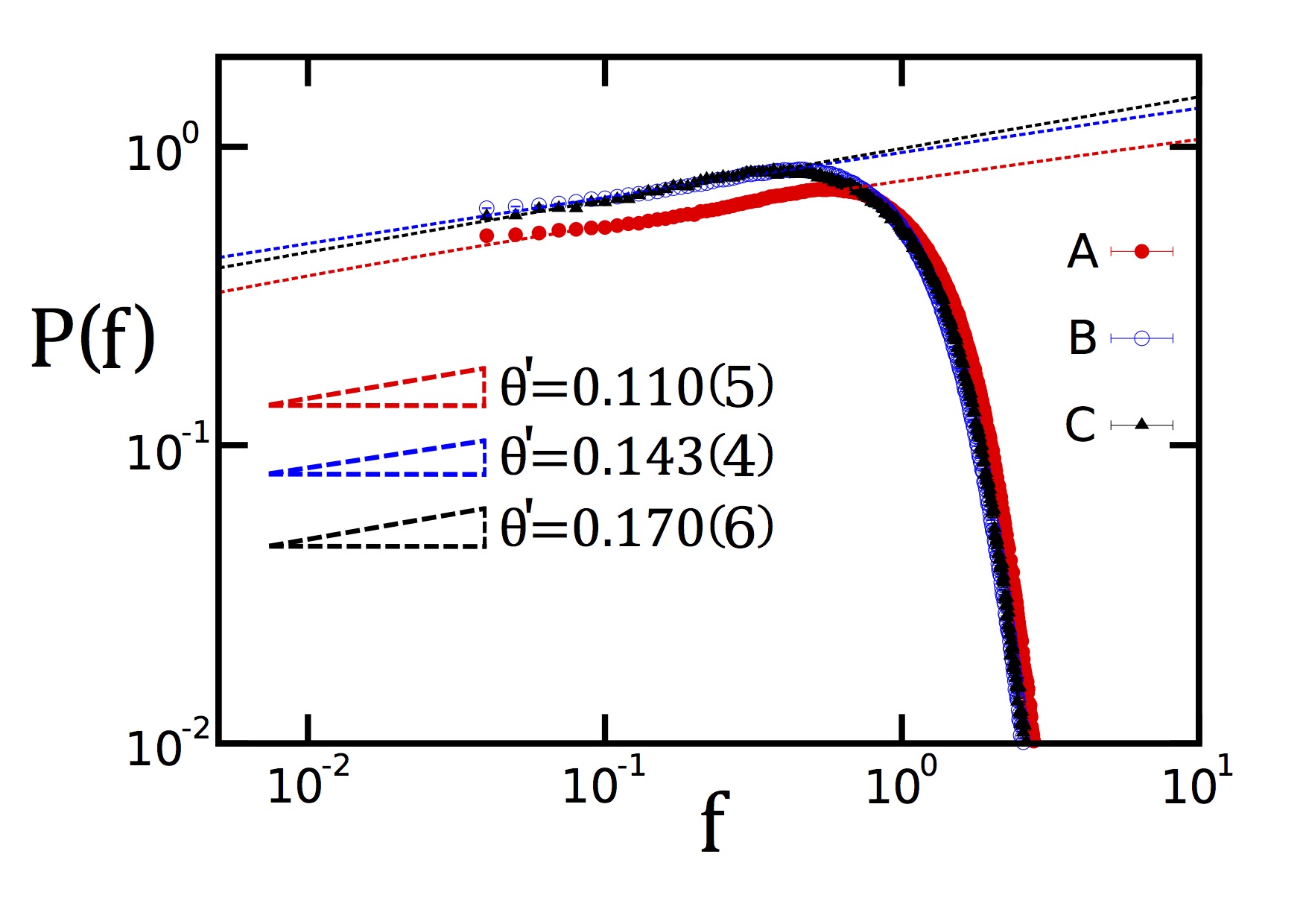}
(b)\includegraphics[width=.5\textwidth]{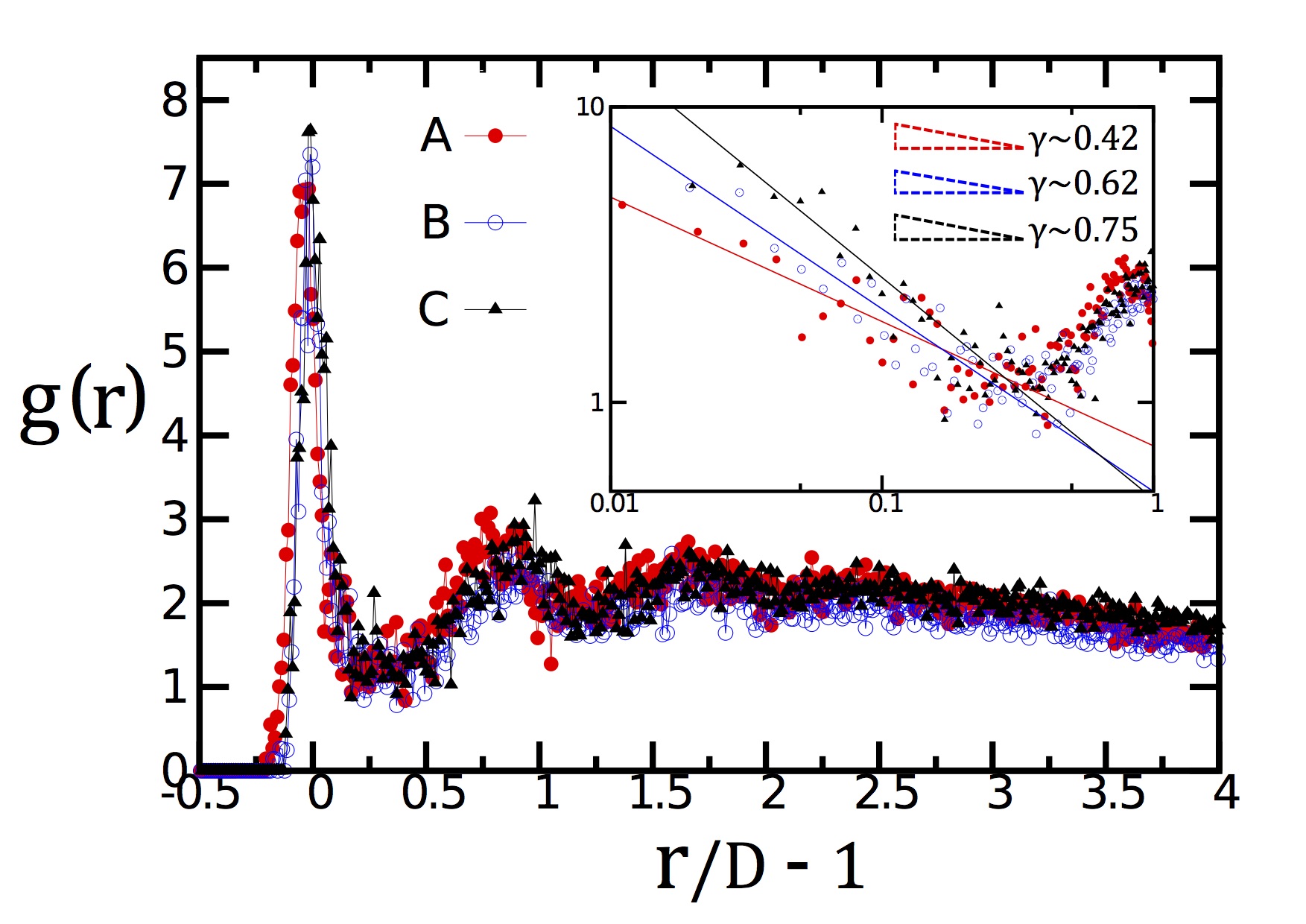}}
\centerline{(c)\includegraphics[width=.5\textwidth]{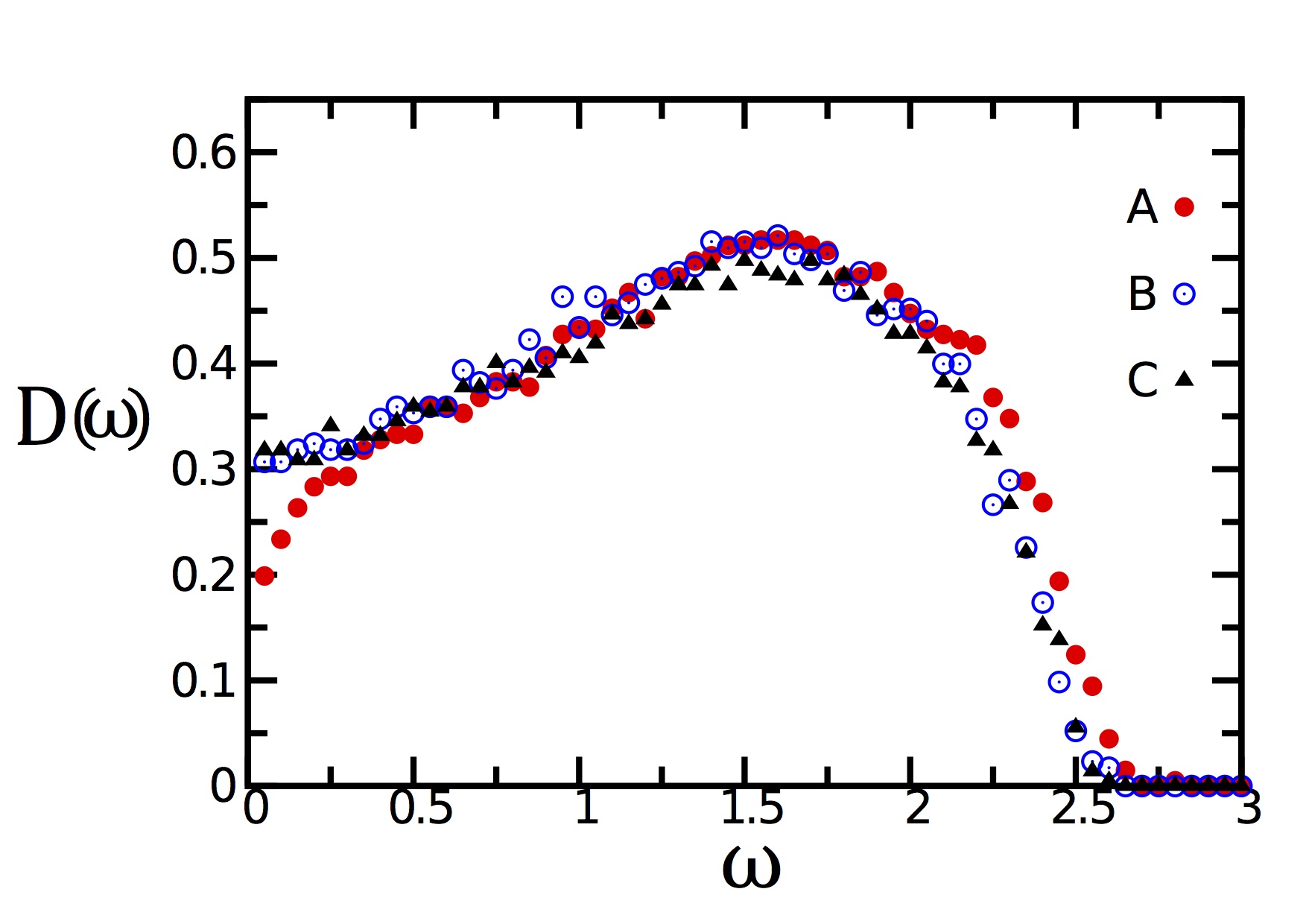}
(d)\includegraphics[width=.5\textwidth]{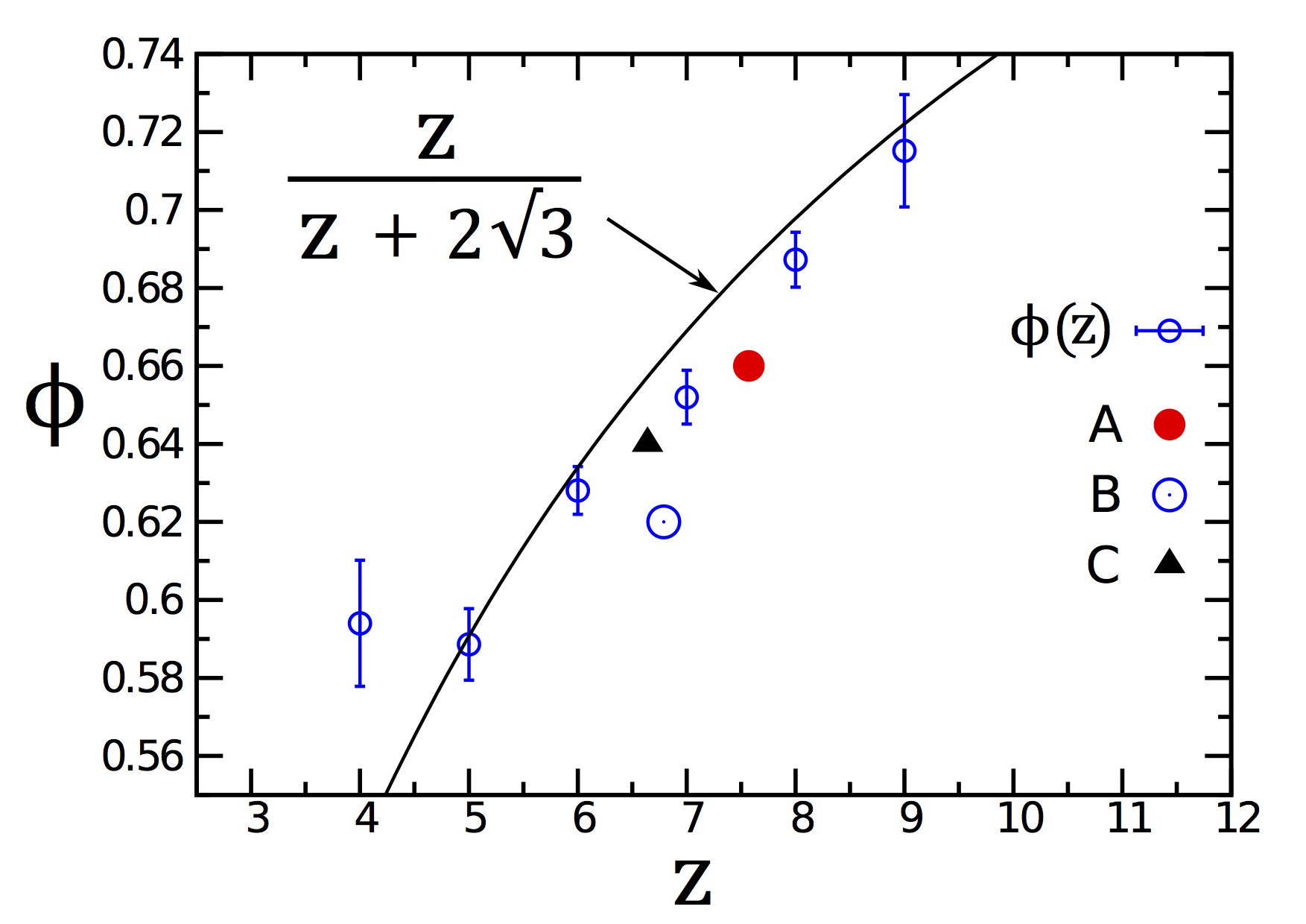}}
\caption{(a) Log-log plot of the inter-particle force 
distribution for packings $A, B$ and $C$ 
(error bars are s.e.m. and are of the same size as the symbols).
Straight lines are best fits for the weak forces with
corresponding scaling exponent $\theta'$.
(b) Main panel: pair correlation functions for
the three packings.  Inset: log-log plot of $g(r)$ 
close to $r\sim
D$. Straight lines are best fits on the small gap
region with corresponding scaling exponent $\gamma$.
(c) Vibrational density of states $D(\omega)$ 
for the three packings. Packings $B$ and $C$ show
an excess of soft modes typical of marginally 
stable jammed packings. Packing $A$ has a different 
behavior in that it shows a deficit of soft modes 
(see discussion in main text). (d) Equation of state for
local per-particle $\phi$ vs local $z$ (blue open circles). 
The black curve is the theoretical prediction~\cite{swm}. 
The three points correspond to the global volume 
fraction and average coordination
number for the three packings.}
\label{fig:results}
\end{figure}

The profiles of the obtained force distributions are shown in
Fig.~\ref{fig:results}a. We obtain values of $\theta'$ ranging from
0.11-0.17 for the three packings (Table \ref{tab:exponents}).  We note
that, in general, the presence of shear jamming can affect the scaling
law of force distribution. In our case, each experimental packing was
prepared by gravitational centrifugation.  This preparation protocol
may generate not only bulk jamming but also shear jamming, and these
modes are responsible for jamming of the packing.

\begin{table}
  \begin{center}
  \caption{$N$ is the number of particles
  whose centers are inside the field of view, average coordination number $\overline{z}$, packing density
  $\phi$, scaling exponent $\theta'$ of the weak force distribution,
  scaling exponent $\gamma$ of the small gap distribution, and
  scaling relation $\gamma >1/(2+\theta')$ \cite{wyart1} for the three
  packings $A, B$ and $C$.}  
  \label{tab:exponents}
    \vspace{1em}
    \begin{tabular}{ccccccc}
      \hline
      \hline
      \hspace{0.2cm} Packing \hspace{0.2cm} &
      \hspace{0.2cm} $N$ \hspace{0.2cm} & 
      \hspace{0.2cm}  $\overline{z}$ \hspace{0.2cm} &
      \hspace{.2cm} $\phi$ & 
      \hspace{.4cm} $\theta'$\hspace{.4cm}  &
      \hspace{.4cm} $\gamma$\hspace{.4cm} &
      \hspace{.4cm} $1/(2+\theta')$ \cite{wyart1}\hspace{.4cm} \\ 
      \hline
      
      \hspace{.6cm}$A$ & 
      \hspace{.2cm} $1393$\hspace{.2cm} &
      \hspace{.2cm} $7.57$\hspace{.2cm} & 
      \hspace{.2cm} $0.66(8)$ \hspace{.2cm}&
      \hspace{.2cm} 0.110(5)\hspace{.2cm} &
      \hspace{.2cm} 0.42(2)\hspace{.2cm} &
      \hspace{.2cm} 0.474(1)\hspace{.2cm} \\ 

     \hspace{.6cm}$B$ &
     \hspace{.2cm} $1263$\hspace{.2cm} & 
     \hspace{.2cm} $6.79$\hspace{.2cm} & 
     \hspace{.2cm} $0.62(4)$ \hspace{.2cm}&
     \hspace{.2cm} 0.143(4)\hspace{.2cm} &
     \hspace{.2cm} 0.62(2)\hspace{.2cm} &
     \hspace{.2cm} 0.467(1)\hspace{.2cm}  \\
    
     \hspace{.6cm}$C$ &
     \hspace{.2cm} $1486$\hspace{.2cm} &
     \hspace{.2cm} $6.64$\hspace{.2cm} &
     \hspace{.2cm} $0.64(7)$ \hspace{.2cm}&
     \hspace{.2cm} 0.170(6)\hspace{.2cm} &
     \hspace{.2cm} 0.75(3)\hspace{.2cm}&
     \hspace{.2cm} 0.461(1)\hspace{.2cm}  \\

     \hline
    \end{tabular}
  \end{center}
\end{table}

The exponent $\theta'$ (as well as $\theta$) is tightly related to
another critical exponent, which controls the behavior of the pair
correlation function $g(r)$ for $r\sim D$, where $D$ is the diameter
of the particles. Specifically, the scaling law \cite{Donev} in 
equation \ref{scalingLaw} holds true for $r\to D$.

\begin{equation}
g(r)\sim (r/D-1)^{-\gamma},
\label{scalingLaw}
\end{equation}

Indeed, the exponents $\gamma$ and $\theta'$ satisfy the inequality
\begin{equation}
\gamma\geq1/(2+\theta'),
\end{equation} 
which is a consequence of the marginal stability of the jammed
packing~\cite{wyart1}.  The values of the exponent $\gamma$ for the
three packings are reported in Table \ref{tab:exponents}, and the
profiles of the $g(r)$ are shown in Fig.~\ref{fig:results}b. With our
values of $\gamma$ and $\theta'$ we find that packings $B$ and $C$
satisfy the theoretical bound, while packing $A$ does not. The reason
is that $A$ turns out to be hyperstatic and, therefore, it is not
supposed to satisfy the bound (we elaborate on this point below).  The
theoretically predicted mean-field value of the
exponent~\cite{charbonneau} is $\gamma=0.41269$, which is outside the
numerical errors of our measured values for packings B and C, and
agrees with packing A.  We notice that the exponents we found for the
force distribution at weak forces, and the pair correlation function
at small gaps do not match the theoretical predictions from the
replica theory of hard sphere glasses.  Possible reasons of this
discrepancy could be the natural particle size polydispersity and
asphericity in the constitutive particle shapes of the experimental
system.  Each packing has roughly 10\% polydispersity and roughly 10\%
of all the particles in each packing have asphericity.  Current
technology (Bang Lab) does not produce perfectly monodisperse silica
particles which are simultaneously amenable to the grafting techniques
required for the present contact detection process.  A numerical study
on how the small polydispersivity in our sample may affect the value
of the exponents would complement the present measurements and test
whether these universal exponents might affected by disorder.

However, marginality can still be suggested beyond the actual value of
the exponents, as long as the exponents satisfy the inequality
$\gamma\geq1/(2+\theta')$, which is a consequence of the marginal
stability of the jammed packing. Thus, this inequality suggests
marginality in packings $B$ and $C$ (in a weak sense), which is indeed
less stringent than perfectly matching with predictions of replica
theory, for which the inequality would be saturated.

We notice that certain expected features of $g(r)$ for
monodisperse packings are not seen in our results. For instance,
there is no clear split-second peak, and the right-hand side of the
split second peak is not at 1D. We think that these features
are not present due to the polydispersivity and asphericity of the
constitutive particles in the packing. We also
notice that the scaling exponent of the week force and the
relationship between $\gamma$ and $\theta'$ have been theoretically
predicted for dry jammed packings, not for colloidal systems in the
presence of hydrodynamic forces. However, we believe that the
presence of the strong centrifugation forces produces a jammed
static packing that may be comparable to theories for dry packings.

The knowledge of the contact network allows us to compute the
vibrational density of states $D(\omega)$ of the packing
~\cite{wyart1,wyart2,vanhecke,silbert}.  This quantity is obtained by
computing the spectrum of the dynamical matrix $\hat{M}$ (and does not
need the existence of a force-law), which is defined as:
\begin{equation}
M_{ij} = -\delta_{\langle ij\rangle}\vec{n}_{ij}\otimes\vec{n}_{ij} + 
\delta_{ij}\sum_{k\in\partial i}\vec{n}_{ik}\otimes\vec{n}_{ik}\ , 
\end{equation}
where the $\vec{n}_{ij}$'s are unit vectors directed from particle $i$
to particle $j$.  The vibrational modes $\omega_i$ are the square root
of the eigenvalues $\lambda_i$ of the matrix $\hat{M}$,
i.e. $\omega_i=\sqrt{\lambda_i}$, and $D(\omega)$ is the probability
density function of those $\omega_i$'s.

Thus, we obtain the vibrational density of states of the packing and the 
concomitant soft modes~\cite{wyart1,wyart2,vanhecke,silbert} 
from the dynamical matrix $\hat{M}$.  In Fig.~\ref{fig:results}c we
show the function $D(\omega)$ for the three packings.  We find that
the density of states for packing $B$ and $C$ shows an excess of low
frequency modes (a flat profile as $\omega\to 0$) as compared to a
crystalline solid, also known as the ``boson peak" in the glass
literature \cite{zamponi1} which is typical of marginally stable
jammed packings. The situation is again different for packing $A$,
which is discussed below.  
We also obtain the equation of state in the
packing fraction-coordination number plane $(\phi, z)$, 
shown in Fig.~\ref{fig:results}d, along with theoretical 
predictions for this quantity\cite{swm}.

\section{D\MakeLowercase{iscussion and conclusions}}

Our results can be interpreted as follows.  Ideally, for an isostatic
packing of hard spheres \cite{baule,torquato,zamponi1} in $3$D, the average
coordination number is $\overline{z}=6$. However, our colloidal
particles are not perfectly spherical.  There is a small fraction of
nonspherical particles some of which are visible in
Fig. \ref{fig:schematic}b, c.  Because asphericity increases
$\overline{z}$, we expect the average coordination to be larger than
six in our jammed colloids.  In packings $B$ and $C$ we find similar
values for $\overline{z}$: $\overline{z}_B=6.79$ and
$\overline{z}_C=6.64$ (see Table \ref{tab:exponents}), and an excess
of low-frequency modes in Fig. \ref{fig:results}.  Therefore, $B$ and
$C$ could reasonably be viewed as being close to marginally stable.
On the contrary, for $A$ we find $\overline{z}_A=7.57$, which is
significantly larger than $\overline{z}_B$ and $\overline{z}_C$.
Moreover, the profile in its vibrational density of states
$D_A(\omega)$ shows a drop at low frequencies (see
Fig.~\ref{fig:results}c), which indicates a possible Debye's
deviation.  Therefore, packing $A$ can be considered as being far from
marginally stable.  This might be caused by excessive compression of
particles that are not perfectly hard due to PEG that is grafted on
the surface of the silica particles.  In fact, in our experiment, the
system is subjected to a uniaxial compression via centrifugation.
This may slightly deform the grafted PEG, thereby closing gaps between
particles and increasing the number of contacts (in comparison to
perfectly hard spheres).

In turn, closing more contacts decreases the amplitude of possible
displacements in particles, which in other words, increases the
frequency of their vibrations. This causes a depletion of the
low-frequency modes with consequent population of the high frequency
modes.  The effect is visible in Fig.~\ref{fig:results}c for
packing A.  Also, the force distribution $P_A(f)$
(see Fig.~\ref{fig:results}a) shows an excess of large forces and a
deficit of small forces compared to $P_B(f)$ and $P_C(f)$.  Again,
this may be caused by compression and
a surplus of contacts in the packing, since, by closing more contacts,
weak forces become more rare, while larger forces become more
probable.  Finally, the pair correlation function $g_A(r)$
(see Fig.~\ref{fig:results}b) is slightly shifted towards the left as
compared to $g_B(r)$ and $g_C(r)$, meaning that smaller gaps are found
with lower probability in packing $A$.  In light of all these
findings, we believe that packing $A$ is hyperstatic.
Incidentally, $A$ is the only packing violating the bound $\gamma\geq
1/(2+\theta')$, which, in fact, is only valid for isostatic packings.

Overall, our experiments suggest that QD fluorophore signal exclusion
proves effective at locating inter-particle contacts in a packing of
colloidal particles.  This contact detection method highlights the
importance of high resolution of inter-particles contacts for testing
theoretical predictions close to the marginally stable state because
these contacts may be used in rigorous computations of force
distribution, yield stress and stability of packings. The fact that
fluorescence (not size) of QD is what is being measured implies that
any fluorophore probe could be used to resolve the inter-particle
space. In particular, smaller probes could be utilized to enhance
resolution beyond that considered here.  Moreover, the present method
could open up a promising experimental field for future studies on
packing of non-spherical particles
\cite{jaeger,solomon,review,baule1,baule2}.

\section*{Acknowledgements}

We acknowledge the support from NSF CMMT (Grant No. DMR-1308235) and
DOE Geosciences Division (Grant No. DE-FG02-03ER15458). We thank
M. Shattuck and W. Rossow for assistance in developing the particle
tracking algorithm, and S. Mukhopadhyay and C. Maldarelli for help
with the experiments.

\section*{Author contributions}
E. K-N. and H. A. M. wrote the manuscript. 
E. K-N. and M. L. G. prepared figures 1a-c.
E. K-N., W. L., and S. L. prepared figures 2a-c and figure 3.
F. M. prepared figures 4a and 4c. E. K-N. prepared figure 4b.
E. K-N. and F. M. prepared figure 4d.
E. K-N. prepared the video file accompanying the manuscript.

All authors reviewed the manuscript.

\section*{Additional Information}

{\bf Competing financial interests:} The authors declare no competing financial interests. 


\clearpage

\begin{thebibliography}{99}

\bibitem{jaeger} Jaeger, H. M. Toward jamming by design. {\textit Soft Matter} {\bf 11}, 12-27 (2015).

\bibitem{solomon} Glotzer, S. C. and Solomon, M. J. 
Anisotropy of building blocks and their assembly into 
complex structures. {\textit Nature}, 
{\bf 6}, 557-562 (2007).

\bibitem{baule} Baule, A., Morone, F., Herrmann, H. J.
and Makse, H. A.
Edwards statistical mechanics for jammed granular matter. 
{\it Rev. Mod. Phys.}
(submitted) arXiv:1602.04369.





\bibitem{edwards2} Edwards, S. F. in 
{\it Granular Matter: An Interdisciplinary Approach},
(ed. Mehta, A.) 121-140 (Springer-Verlag, 1994).

\bibitem{vanhecke} Snoeijer,  J. H., Vlugt, 
T. J. H., van Hecke,  M. and van Saarloos, W. 
Force network ensemble: a new approach to static
granular matter.
{\it Phys. Rev. Lett.}, {\bf 92}, 054302 (2004).


\bibitem{forcemap} Henkes, S. and Chakraborty, B. 
Jamming as a critical phenomenon: a field theory of 
zero-Temperature grain packings.
{\it Phys. Rev. Lett.}, {\bf 95}, 198002 (2005).

\bibitem{kurchan} Krzakala, F. and Kurchan, J.  
Landscape analysis of constraint satisfaction 
problems.
{\it Phys. Rev. E}, {\bf 76}, 021122 (2007).

\bibitem{swm} Song, C., Wang, P. and Makse, H. A.
A phase diagram for jammed matter.
{\it Nature}, {\bf 453}, 629 (2008).

\bibitem{torquato} Torquato, S. and Stillinger, F. H.
Jammed hard-particle packings: From Kepler to 
Bernal and beyond.
{\it Rev. Mod. Phys.}, {\bf 82}, 2633
(2010).

\bibitem{zamponi1} Parisi, G. and Zamponi, F. 
Mean field theory of hard sphere glasses and jamming.
{\it Rev. Mod. Phys.}, {\bf 82}, 789 (2010).


\bibitem{wyart1} Wyart, M. 
Marginal stability constrains pair and 
force distributions at random close packing.
{\it Phys. Rev. Lett.}, {\bf 109}, 125502 (2012).

\bibitem{wyart2} Lerner, E., D\"{u}ring, G. and Wyart, M. 
Low-energy non-linear excitations in sphere packings.
{\it Soft Matter}, {\bf 9}, 8252-8263 (2013).

\bibitem{charbonneau} Charbonneau, P., Kurchan, J.,
Parisi, G., Urbani, P. and Zamponi, F. 
Fractal free energy landscapes in structural glasses.
{\it Nature Comm.}, {\bf 5}, 3725 (2014).

\bibitem{aste1} Aste, T., Saadatfar, M. and Senden, T. J.
Geometrical structure of disordered sphere packings.
{\it Phys. Rev. E}, {\bf 71}, 061302 (2005).

\bibitem{aste2} Schaller, F. M., Neudecker, M., Saadatfar,
M., Delaney, G., Mecke,  K., Schr{\" o}eder-Turk, G. E. 
and Schr{\" o}ter, M. 
Tomographic analysis of jammed ellipsoid packings.
{\it AIP Conf. Proc.}, {\bf 1542}, 377 (2013).



\bibitem{weeks2} Weeks, E. R., Crocker, J. C., Levitt, A.C. 
and Schofield, A. 
Three-dimensional direct imaging of structural 
relaxation near the colloidal glass transition.
{\it Science}, {\bf 28}, 627 (2000).

\bibitem{brujic} Bruji{\'c}, J., Edwards, S. F.,
Grinev, D. V., Hopkinson, I., Bruji{\'c}, D.
and Makse, H. A. 
3D bulk measurements of the force distribution in a
compressed emulsion system.
{\it Faraday Discuss.}, {\bf 123}, 207 (2003).

\bibitem{jorjadze} Jorjadze, I.,  Pontani, L-L., 
Newhall, A., and Bruji{\'c}, J.
Attractive emulsion droplets probe the phase 
diagram of jammed granular matter.
{\it Proc. Nat. Acad. Sci. U.S.A.}, {\bf 108}, 4286 (2011).

\bibitem{dinsmore} Zhou, J., Long, S., Wang,  Q. 
and Dinsmore, A. D.
Measurement of forces inside a three-dimensional 
pile of frictionless droplets.
{\it Science}, {\bf 312}, 1631 (2006).

\bibitem{brujic2} Bruji{\'c}, J., Song, C.,
Wang, P., Briscoe, C., Marty, G. and Makse, H. A. 
Measuring the Coordination Number and Entropy 
of a 3D jammed emulsion packing by confocal microscopy.
{\it Phys. Rev.Lett.}, {\bf 98}, 248001 (2007).

\bibitem{clusel} Clusel, M., Corwin, E. I., 
Siemens, A. O. N. and Bruji{\'c}, J.
A `granocentric' model for random packing of 
jammed emulsions.
{\it Nature}, {\bf 460}, 611 (2009).

\bibitem{NyombiThesis} Kyeyune-Nyombi, E. 
{\it Experimental
Studies of Jamming in Colloidal Systems using 
Fluorescence Microscopy, Ph. D. thesis} 
(City College of New York, 2016).

\bibitem{Hermanson} Hermanson, G. T. {\it Bioconjugation Techniques, 2nd ed}. (Elsevier Inc., 2008).

\bibitem{Israelachvilli} Israelachvilli, J. N. 
{\it Intermolecular and Surface Forces, 3rd ed}.
(Elsevier Inc., 2011).

\bibitem{Drobek} Drobek, T. and Spencer, N. Nanotirbology 
of surface-grafted PEG layers in an aqueous environment.
{\it Langmuir}, {\bf 24}, $1484-1488$ (2008).

\bibitem{Giepmans} Giepmans, B. N., Deerinck, T. J., 
Smarr,  B. L., Jones,  Y. Z. and Ellisman, M. H. 
Correlated light and electron microscopic 
imaging of multiple endogenous proteins using 
quantum dots.
{\it Nature}, {\bf 2}, 743 (2005).


\bibitem{Asakura} Asakura, S. and Oosawa, F. 
On interaction between two Bodies immersed in 
a solution of macromolecules.
{\it J. of Chem. Phys.}, {\bf 22}, 1255 (1954). 

\bibitem{Hell} Hell, S. W. and Wichman, J.
Breaking the diffraction resolution limit by
stimulated emission: stimulated-emission-depletion
fluorescence microscopy.
{\it Opt. Lett.}, {\bf 19}, 780 (1994).

\bibitem{Betzig2} Betzig, E., Patterson, G. H., 
Sougrat, R., Lindwasser, O. W., Olenych, S.,
Bonifacino, J. S., Davidson, 
M. W., Lippincott-Schwartz,  J. and Hess, H. F.
Imaging intracellular fluorescent proteins at 
nanometer resolution.
{\it Science}, {\bf 313}, 1642 (2006).



\bibitem{mari} Mari, R., Krzakala, F. and Kurchan, J.
Jamming versus glass transitions.
{\it Phys. Rev. Lett.}, {\bf 103}, 025701 (2009).

\bibitem{degiuli} DeGiuli, E., Lerner, E. and Wyart, M. 
Force distribution affects vibrational properties 
in hard-sphere glasses.
{\it Proc. Nat. Acad. Sci. U.S.A.}, 
{\bf 111}, 17054 (2014).

\bibitem{linbo} Bo, L., Mari, R., Song, C. and Makse, H. A.
Cavity method for force transmission in jammed 
disordered packings of hard particles.
{\it Soft Matter}, {\bf 10}, 7379-7392 (2014).

\bibitem{lin} Lin, J., Jorjadze, I., Pontani, L-L., 
Wyart, M., and Bruji{\'c}, J.
Evidence for marginal stability  in emulsions.
{\it Phys. Rev. Lett.}, {\bf 117}, 208001 (2016).

\bibitem{buckling} Charbonneau, P., Corwin, E. I., 
Parisi, G. and Zamponi, F. 
Jamming criticality revealed by removing 
localized buckling excitations.
{\it Phys. Rev. Lett.}, {\bf 114}, 125504 (2015).

\bibitem{Donev} Donev, A., Torquato, S. 
and Stilinger, F. H. 
Pair correlation function characteristics of nearly
jammed disordered and ordered hard-sphere packings.
{\it Phys. Rev. E}, {\bf 71}, 011105 (2005).

\bibitem{silbert} Silbert, L. E., Liu,  A. J. and 
Nagel, S. R. 
Vibrations and diverging length scales near 
the unjamming Transition.
{\it Phys. Rev. Lett.}, {\bf 95}, 098301 (2005).

\bibitem{review} Makse, H. A., Bruji\'c, J.
and Edwards, S. F. 
Statistical Mechanics of Jammed Matter, in 
{\it The Physics of Granular Media}
(edited by Hinrichsen H. and Wolf, D. E.)
(Wiley-VCH, Weinheim, 2004).


\bibitem{baule1} Baule, A., Mari, R., Bo, L., 
Portal, and Makse, H. A.
Mean-field theory for random close packings 
of axisymmetric particles.
{\it Nature Comm.} {\bf 4}, 2194 (2013).


\bibitem{baule2} Baule, A. and Makse, H. A. 
Fundamental challenges in packing problems: 
from spherical to non-spherical particles. 
{\it Soft Matter} {\bf 10}, 4423-4429 (2014).

\newcounter{firstbib}
\setcounter{firstbib}{\value{NAT@ctr}}
\end{thebibliography}
\end{document}